\documentclass[12pt]{article}

\usepackage{amsmath,amssymb}
\usepackage{bm}
\usepackage{graphicx} 
\usepackage{subfigure}

\makeatletter

\@addtoreset{equation}{section}
\makeatother

\newcommand{\beq}{\begin{equation}}
\newcommand{\eeq}{\end{equation}}

\begin{document}

\baselineskip=18pt  
\baselineskip 0.8cm

\begin{titlepage}

\setcounter{page}{0}

\renewcommand{\thefootnote}{\fnsymbol{footnote}}

\begin{flushright}
{{\tt CALT-68-2796}\\
{\tt IPMU10-0110}}
\end{flushright}

\vskip 1cm

\begin{center}
{\Large \bf
Holographic End-Point of \\ Spatially Modulated Phase Transition}

\vskip 2cm

{\large Hirosi Ooguri and Chang-Soon Park}

\vskip 1cm

{
\it
California Institute of Technology, Pasadena, CA 91125, USA,\\
and\\
Institute for the Physics and Mathematics of the Universe,\\
 University of Tokyo, Kashiwa 277-8586, Japan\\}

\end{center}

\vspace{1cm}

\centerline{{\bf Abstract}}
\medskip
\noindent
In the previous paper [arXiv:0911.0679], we showed that 
the Reissner-Nordstr\"om black hole in the 5-dimensional anti-de Sitter 
space coupled to the Maxwell theory with the Chern-Simons term 
is unstable when the Chern-Simons coupling is sufficiently large.
In the dual conformal field theory, the instability suggests
a spatially modulated phase transition. 
In this paper, we construct and analyze non-linear solutions which describe
the end-point of this phase transition. In the limit where the 
Chern-Simons coupling is large, we find that 
the phase transition is of the second order 
with the mean field critical exponent. However, the 
dispersion relation with the Van Hove singularity
enhances quantum corrections in the bulk,
and we argue that this changes the order of the phase transition from 
the second to the first. We compute linear response functions in the 
non-linear solution and find an infinite off-diagonal DC conductivity
in the new phase.

\end{titlepage}
\setcounter{page}{1} 


\section{Introduction}

In the previous paper \cite{Nakamura:2009tf}, together with Shin Nakamura, 
we pointed out that the Maxwell theory with the Chern-Simons term in 
the 5-dimensional Minkowski space is tachyonic when a constant
electric field is turned on. A similar mechanism leads
to an instability of charged black hole in the 5-dimensional 
anti-de Sitter space ($AdS_5$) if the Chern-Simons coupling 
for the Maxwell field is sufficiently large. 
Interestingly, the instability modes carry non-zero momenta along 
the boundary of $AdS_5$. This suggests that there is a novel phase 
transition in the holographically dual field theory
 at finite chemical potential, where order parameters acquire 
spatially modulated expectation values. 

The analysis of our previous paper was at the linearized level, and what
we observed was an onset of the phase transition. To understand the nature
of the new phase which emerges as the end-point of the instability, 
we need to examine full non-linear solutions to the equations of motion
in the bulk. In this paper, we construct such 
solutions in the limit where the Chern-Simons coupling is large
and back-reaction of the Maxwell field to the metric is negligible. 
This is analogous to the probe limit employed in \cite{Hartnoll:2008vx}. 
Using the solutions, we compute
the expectation values of the order parameters near the phase transition 
temperature and find that the phase transition is of the 
second order with the mean field critical exponent. 

The Chern-Simons term modifies the dispersion relation 
in such a way that the density of states per unit 
energy diverges at some non-zero momenta, causing the Van Hove singularity.
Moreover, this happens even above the phase transition temperature. 
It suggests that quantum corrections
to the phase transition can be significant. 
We argue that the order of the phase transition is changed from the second 
to the first due to quantum effects in the bulk. 

The rest of the paper is organized as follows. In section 2, 
as a warm-up exercise, we will discuss non-linear solutions to
the Maxwell-Chern-Simons theory in the 5-dimensional Minkowski
space. In section 3, we turn to the theory in the full black hole geometry and
construct non-linear solutions in the limit where the Chern-Simons coupling
is large.
We find that the phase transition is of the second order with the
mean field exponent. In section 4, we discuss quantum corrections to the
phase transition and argue the order of the phase transition 
is changed.  We evaluate the linear response of the system 
in section 5.
In appendix A, we discuss non-linear solutions in $AdS_2\times \mathbb{R}^3$, which is the near 
horizon limit of the extremal black hole.

\noindent Note added: After the first version of this paper was completed, 
we were informed of the work \cite{Sachdev:2010um}, which suggested
that an instability to crystalline phases might be a generic feature 
of phases which are describable by a bulk $AdS_2$ geometry. 
Such an instability would
provide a natural way to understand the ground state entropy.

\section{Maxwell-Chern-Simons Theory in Minkowski Space}\label{sec:MCSflat}

In this section, we consider the Maxwell theory with the Chern-Simons term
in the 5-dimensional Minkowski space.
The Lagrangian is given by
\begin{equation}
\mathcal{L}=-\frac 1 4 F_{IJ} F^{IJ} + \frac{\alpha}{3!} \epsilon^{IJKLM} A_I F_{JK} F_{LM}\;,
\end{equation}
where $I,J,\ldots$ run from 0 to 4.
The equations of motion are
\begin{equation}
\partial_J F^{JI} + \frac{\alpha} 2 \epsilon^{IJKLM} F_{JK} F_{LM} =0\;.
\label{eom}
\end{equation}
In particular, the time component of the above 
can be written as the Gauss law,
\begin{equation}\label{E:GaussLawFlat}
\partial_A \Pi^A =0\;,\qquad \text{where $\Pi^A = -F^{A0} + \alpha \epsilon^{ABCD} A_B F_{CD}$}\;,
\end{equation}
where the indices $A,B,\ldots$ run from 1 to 4.

A constant electric field is a solution to the equations 
of motion. However, as shown in \cite{Nakamura:2009tf}, there are 
unstable modes in the following range of momentum,
\begin{equation}
0 < |\vec{k}| < 4\alpha E,
\label{inequality}
\end{equation}
where $E$ is the background electric field
and $\vec{k}$ is a projection of the spatial momentum 
onto the plane orthogonal the electric field.  
Let us describe the instability mode found in \cite{Nakamura:2009tf}.
If the electric field is in the $x^1$ direction, it is convenient
to decompose the 5-dimensional momentum as 
$(p_{\mu}, k_i)$, $\mu=0,1$ and $i=2,3,4$. 
Consider a linear fluctuation of the Maxwell field of the
form, 
\begin{equation}
a_i = c_i^{(\pm)} e^{i p_{\mu} x^{\mu} + i k_i x^i},
\label{ansatz}
\end{equation}
where 
$c_i^{(\pm)}$ are an eigenvectors of $\epsilon_{ijk} k_j$,
\begin{equation}
  \epsilon_{ijk} k_j c_k^{(\pm)} = \pm |\vec k| c_i^{(\pm)},
\end{equation}
obeying the transverse gauge condition $k_i c_i^{(\pm)}=0$. 
Substituting (\ref{ansatz}) into the equations of motion (\ref{eom}), 
we find the dispersion relation for this mode as 
\begin{equation}
(p_0)^2-(p_1)^2 = k^2 \mp 4\alpha E k\ ,~~~k =|\vec{k}|.
\end{equation}
This means that it is tachyonic for the range (\ref{inequality})
if we choose $c_i^{(+)}$.

\subsection{Non-linear Solutions}

We can find a non-linear solution triggered by the perturbation
$a_i = c_i^{(+)} e^{i p_{\mu} x^{\mu} + i k_i x^i}$ as follows. 
Although the unstable mode breaks the translational invariance 
along the direction
of the momentum $\vec{k}$, it is invariant under some
combination of translation along $\vec{k}$ and rotation in
the transverse plane. The solution is also translationally
invariant in the transverse directions. It is then natural 
to look for a non-linear solution
with the same set of symmetries, and we choose the following ansatz,
\begin{equation}\label{E:FlatAnsatz}
\begin{split}
A_0&=0\;,\qquad A_2=g(t)\\
A_1&=A_1(t)\\
A_3+ i A_4 &= h(t) e^{-i k x_2}\;.
\end{split}
\end{equation}
We denote the time coordinate by $x^0$ or $t$ interchangeably.
The equation of motion for the function $g(t)$ sets $g(t)=0$.
The remaining equations of motion become
\begin{equation}\label{E:EOMflat5D}
\begin{split}
&\ddot{A}_1(t) -4 \alpha k h(t) \dot h(t)=0\\
&\ddot h(t) + k^2 h(t) + 4\alpha k \dot A_1(t) h(t) =0\;.
\end{split}
\end{equation}
Note that the momentum $\Pi_1$ conjugate to $A_1$ is given by
\begin{equation}\label{E:Pi1Flat}
\Pi_1 = -\dot A_1(t) +2\alpha k h(t)^2 \;.
\end{equation}
The first equation of \eqref{E:EOMflat5D} can be written as 
$\partial \Pi_1/\partial t =0$ and solved by 
$\Pi_1=constant$.
The integration constant is fixed as $\Pi_1 = E$
by the initial configuration
where $h=0$ and  $\Phi_1= -\dot A_1 =E$.

\begin{figure}[ht]
\centering
\includegraphics[width=5cm]{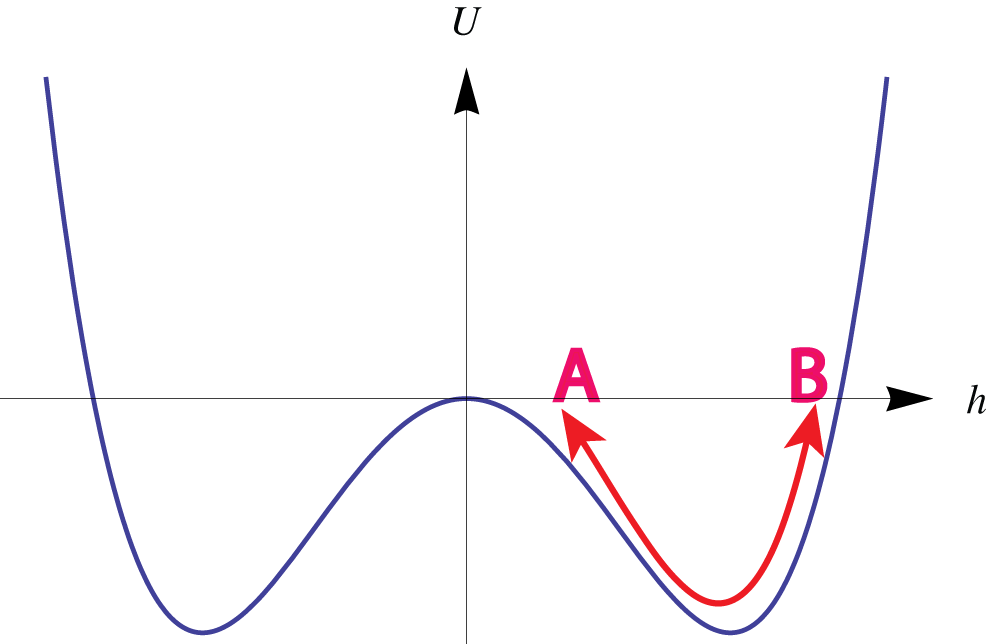}
\caption{Double well potential $U$ for a classical particle with coordinate $h$. If the particle starts slightly outside of the origin, say at $A$, then the particle will oscillate between $A$ and another point $B$ with
the same potential energy.}\label{fig:doublewell}
\end{figure}

From \eqref{E:EOMflat5D} and \eqref{E:Pi1Flat},
\begin{equation}
\ddot h(t) + k^2 h(t) - 4\alpha k E h(t) + 8 \alpha^2 k^2 h^3(t)=0\;.
\end{equation}
This can be integrated and yields
\begin{equation}
\frac 1 2 \dot h(t)^2 - \frac 1 2 k (4\alpha E - k) h(t)^2 + 2\alpha^2 k^2 h(t)^4 = constant.
\end{equation}
This is equal to the energy density $\mathcal{H}$ of the electro-magnetic field,
\begin{equation}
\mathcal{H} = \frac 1 2 F_{0A} F_{0A} + \frac 1 4 F_{AB} F_{AB}\;,
\end{equation}
minus the energy density $\frac{1}{2} E^2$ of the constant electric field. 
Thus, we are effectively considering a classical particle with coordinate $h(t)$ moving 
in the potential
\begin{equation}
U= -\frac 1 2 k(4\alpha E - k)h^2 + 2\alpha^2 k^2 h^4\;.
\label{potential}
\end{equation}
This is a double well potential for $0<k<4\alpha E$ as in 
Figure \ref{fig:doublewell}.
The original homogeneous phase corresponds to the point $h=0$, 
which is unstable.
If we add some perturbation, the amplitude $h(t)$ starts oscillating 
as in the figure.

\subsection{The Final Configuration}

We have seen that the instability induces an oscillatory 
solution in the potential (\ref{potential}). 
Suppose that our system is weakly coupled to a
heat reservoir with a large heat capacity at very low temperature.
Eventually the oscillation will fade away by transferring its energy to
the heat reservoir and the system will land on its lowest energy state.
Let us try to find out the final configuration of this process. 

\begin{figure}[ht]
\centering
\subfigure{
\includegraphics[width=6cm]{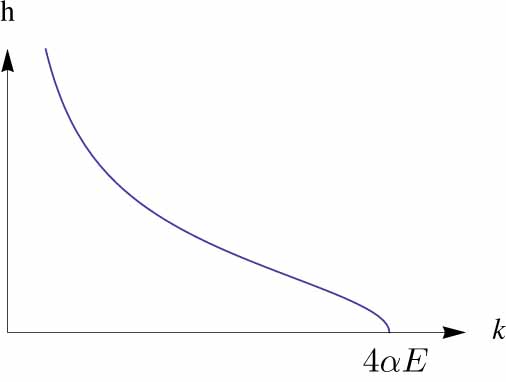}
}
\centering
\subfigure{
\includegraphics[width=6cm]{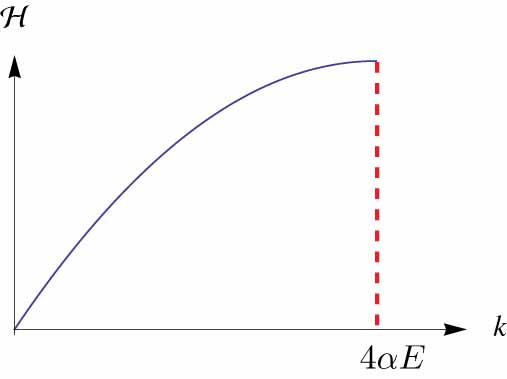}
}
\caption{The amplitude $h$ and the energy density $\mathcal H$ as a function of $k$.}
\label{fig:flatEh}
\end{figure}

For the static solutions, 
\begin{equation}
h = \pm  \frac{1}{2\alpha}\sqrt{\frac{4\alpha E - k}{2k}},
\label{staticsolution}
\end{equation}
which stay at one  of the two minima of the potential, 
the energy density $\mathcal H$ is given by
\begin{equation}
\mathcal H = U + \frac{1}{2} E^2 = 
\frac k {4\alpha} \left( E - \frac{k}{8\alpha}\right)\;.
\end{equation}
Especially, the energy density vanishes when $k=0$.
Figure \ref{fig:flatEh} shows 
the amplitude $h$ and the energy density $\mathcal H$ change as functions of $k$.
Since the energy density is monotonically increasing in $k$,
we expect that solutions with $k\neq 0$ are unstable and decay to 
the solution with $k=0$.
Note that, although $h$ diverges as $k$ goes to $0$, the field strength $E$ 
vanishes in this limit. The constant electric field in the initial
configuration is wiped out in the $k=0$ solution and  the final configuration
will be the trivial vacuum state with $E=0$. 

We can directly check that the static solution (\ref{staticsolution}) with $k\neq 0$ is
unstable. The solution corresponds to the gauge field configuration, 
$A_1^{(0)}(t) = \frac{k}{4\alpha} t$ and
 $A_3^{(0)} + i A_4^{(0)} = \frac{1}{2\alpha}
\sqrt{\frac{4\alpha E - k}{ 2k}} e^{-ik x_2}$.
Let us add a small perturbation $A^{(0)} \rightarrow A^{(0)} + a$
to this background.
Assuming that the modes depend only on $t$ and $x_2$ with $a_0=0$, 
the equations of motion become
\begin{equation}\
\begin{split}
\Box a_1 + 2\alpha k h ( e^{i k x_2} \partial_0 a_+ + e^{-i k x_2} \partial_0 \bar a_+ ) &=0\\
\Box a_2 &=0\\
\Box a_+ + i k \partial_2 a_+ - 4\alpha k h e^{-i k x_2} \partial_0 a_1 &=0\;,
\end{split}
\end{equation}
where $\Box = -\partial_0^2 + \partial_2^2$ and $a_+= a_3 + i a_4$.
The coefficients of the equations have $x_2$ dependence, which can be removed by using 
two real variables $\tilde a_3$ and $\tilde a_4$ such 
that $a_+= e^{-i k x_2} (\tilde a_3 + i \tilde a_4)$ in place of $a$.
Assuming the $t$ and $x_2$ dependence of the fields $a_1$, $\tilde a_3$ 
and $\tilde a_4$ to be of the form $e^{-i \omega t + i q x_2}$ and that $a_2=0$, 
a non-trivial solution exists if and only if
\begin{equation}
(\omega^2 - q^2)\left[\omega^4 - 2\omega^2 (q^2 + k (4\alpha E- k)) + q^2(q^2 - k ^2)\right]=0\;,
\end{equation}
From the second factor, we see that the product of
two solutions for $\omega^2$ is $q^2(q^2 - k^2)$, which is negative for $0<q<k$.
Thus one of the two roots of $\omega^2$ must be negative, 
representing an unstable mode. 
Since the momentum $q$ of the instability mode is in the range $0 < q < k$,
we expect that the solution decays toward the lowest energy 
state with momentum $k=0$. 

We have also performed numerical analysis of time dependent solutions 
with the initial configuration of constant electric field $E$ assuming that solutions depend only on the coordinates $t$ and $x_2$.
We found that a small localized perturbation generates a domain where the electric and magnetic
fields fluctuate and that the domain expands at the speed of light.
The magnetic field in the domain carries a range of momenta, which tend to move to zero momentum state.
The strength of the electric field decays as the domain expands, suggesting that the
system will eventually settle down to the trivial state with $E=0$. 

To summarize, the instability of the constant electric field in 
the Maxwell-Chern-Simons theory in the 5-dimensional Minkowski space, 
which we found in our previous paper 
\cite{Nakamura:2009tf}, leads to the trivial 
vacuum state with no background field strength $F_{IJ}=0$. This reminds 
us of the Schwinger mechanism where a constant electric field is screened 
by virtual production of electron-positron pairs. 

This result should be contrasted with the corresponding instability of
the charged black hole in $AdS_5$, which we will study in the next section.
There, we do not expect that the background electric field to disappear since 
the electric charge of the black hole is fixed by the chemical potential
at the boundary. 
Indeed, we will find stable solutions with non-zero momentum $k$ in this case. 

\section{Maxwell-Chern-Simons Theory in the $AdS_5$ Black Hole Geometry}\label{sec:MCSSBHAdS}

In this section, we will construct non-linear solutions which describe
the end-point of the instability of the charged black hole in $AdS_5$. 
Since the Maxwell field contributes to the energy momentum tensor, in general
we need to analyze the coupled Einstein and Maxwell equations. 
Here we will simplify the problem by taking a limit where we can ignore the back-reaction 
of the Maxwell field to the metric. 

The Lagrangian density for the Maxwell-Chern-Simons theory is given by
\begin{equation}
\mathcal{L} = -\frac 1 4 F_{IJ}F^{IJ} + \frac{\alpha}{3!\sqrt{-g}} \epsilon^{IJKLM} A_I F_{JK} F_{LM}\;.
\end{equation}
Rescaling the gauge field as $\tilde A = \alpha A$,
the Lagrangian density becomes
\begin{equation}\label{E:rescaledLagrangian}
\mathcal L = \frac{1}{\alpha^2} \left(-\frac 1 4 \tilde F_{IJ}\tilde F^{IJ} + \frac{1}{3!\sqrt{-g}} \epsilon^{IJKLM} \tilde A_I \tilde F_{JK} \tilde F_{LM}\right)\;.
\end{equation}
When $\alpha$ is large, for a solution with finite $\tilde A$,
the energy-momentum tensor is of the order $O(1/\alpha^2)$ and 
the coupling of the Maxwell field to the metric is suppressed. 
This limit is analogous to the infinite charge limit considered in the holographic description 
of superconductivity \cite{Hartnoll:2008vx}.

\subsection{The Large $\alpha$ Limit}

To keep $\tilde A$ finite, we have to scale the background gauge field as well. 
This means that the chemical potential $\mu$
of the black hole should also be scaled in such a way
 that the combination $\alpha \mu$ remains finite. 
Let us examine what this limit means to the black hole solution. 
The Reissner-Nordstr\"om solution has the metric
\begin{equation}\label{E:RNblackhole}
ds^2=- H(r) dt^2 + \frac 1 {H(r)} dr^2 + r^2 d{\vec x}^2\;, \qquad \vec x = 
(x^2, x^3, x^4) \;,
\end{equation}
where the function $H(r)$ is given by
\begin{equation}\label{E:RNblackhole2}
H(r)=r^2\left[1-\left(\frac {r_+} r\right)^4\right]\;,
\end{equation}
The temperature in this limit of $\mu \rightarrow 0$ becomes 
\begin{equation}
T=\frac{r_+}{2\pi} \left(2- \frac{\mu^2}{3r_+^2}\right)\rightarrow \frac{r_+}{\pi}\;.
\end{equation}
The background geometry in this limit
is simply the (neutral) Schwarzschild $AdS_5$ solution.
In terms of the rescaled finite gauge field $\tilde{A}$, 
the background field strength is given by
\begin{equation}
\tilde F=\frac{\tilde E}{r^3} dt\wedge dr\;, \qquad \text{where $\tilde E=-2\alpha \mu r_+^2=-\frac{2 r_+^3}{\pi}\frac 1 {\tau}$,}
\end{equation}
where we introduce a new variable $\tau = r_+/\pi \alpha \mu$ 
for later convenience. 
Since $\tau=T/{\mu\alpha}$, it can be thought of as a rescaled temperature.

\begin{figure}[ht]
\centering
\includegraphics[width=7cm]{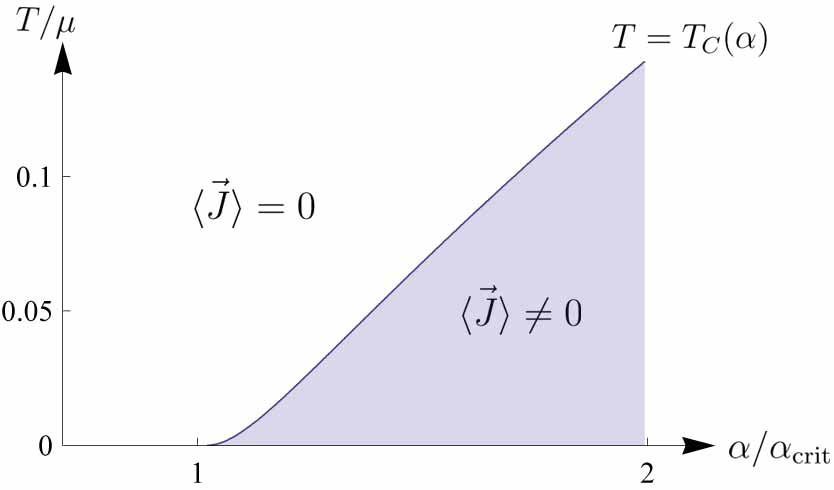}
\caption{Critical temperature as a function of the Chern-Simons coupling $\alpha$. 
The shaded region indicates a phase with a non-zero expectation value of the 
conserved current $\vec{J}$ which is helical and position dependent.}\label{fig:criticaltemperature}
\end{figure}

It is important to note that we have access to the phase transition point 
in this limit. 
In \cite{Nakamura:2009tf}, we studied the instability of 
the Reissner-Nordstr\"om solution
and obtained the critical temperature $T_c$ for the instability as a function of the Chern-Simons coupling $\alpha$.
The result of our numerical analysis is reproduced in Figure \ref{fig:criticaltemperature}.
For large $\alpha$, the dimensionless combination $T_c/\mu$ grows linearly in $\alpha$. 
Thus, we can analyze the behavior of the system near $T=T_c$ by taking the limit of $\alpha \rightarrow
\infty$ while keeping the combination $\alpha \mu$ finite. 

Let us find a non-linear solution to the equations of motion that
describes the spatially modulated phase in this limit.
We look for a solution that has the same symmetry as 
that of the unstable modes found in \cite{Nakamura:2009tf},
namely a linear combination of a translation along $x_2$ and 
a rotation in the 3-4 plane, as well as the translation symmetries 
along $t$, $x_3$ and $x_4$.
This leads to the following ansatz, 
\begin{equation}\label{E:inhomogeneousAnsatz}
\begin{split}
&\tilde A_0 = f(r)\;,\qquad \tilde A_1=g_1(r)\;, \qquad \tilde A_2=g_2(r)\\
&\tilde A:=\tilde A_3 +i \tilde A_4 = h(r) e^{-i k x_2}\;.
\end{split}
\end{equation}
Note that $g_1(r)$ can be set to vanish by a gauge choice, and 
$g_2(r)$ has to vanish by 
the Maxwell equation $\frac{\delta S}{\delta \tilde A_2}=0$.
The non-trivial equations of motion are
\begin{equation}
\begin{split}
-\partial_r (r^3 f'(r))-4 k h(r) h'(r) &=0\\
\partial_r( H(r) r h'(r) ) - \frac 1 r k^2 h(r) + 4 k h(r) f'(r)&=0\;.
\end{split}
\end{equation}
The first equation can be integrated and becomes
\begin{equation}
r^3 f'(r) + 2 k h(r)^2 = \tilde E\;.
\end{equation}
Eliminating $f'(r)$ in the second equation by using the above relation, we obtain
\begin{equation}\label{E:EOMinRNAdSBH}
\partial_r( H(r) r h'(r) ) - \frac 1 r k^2 h(r) + 4 k \frac{h(r)}{r^3} (\tilde E-2 k h(r)^2)=0\;.
\end{equation}

\subsection{Second Order Phase Transition}

\begin{figure}[ht]
\centering
\subfigure{
\includegraphics[width=6cm]{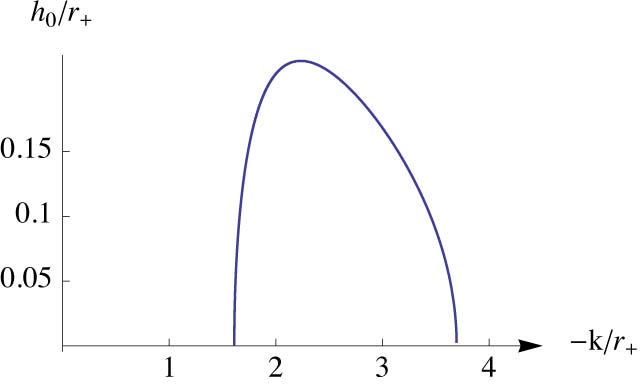}
}
\centering
\subfigure{
\includegraphics[width=6cm]{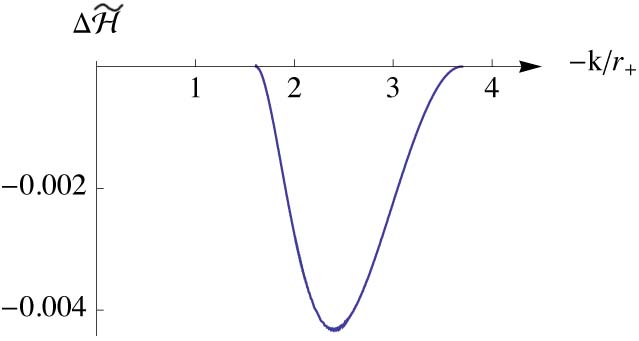}
}
\caption{The amplitude $h_0$ and the energy density difference $\Delta \widetilde{\mathcal H}$ 
from the homogeneous phase as functions of $k$.}
\label{fig:ads5Eh}
\end{figure}

We have solved the differential equation \eqref{E:EOMinRNAdSBH} numerically.
For each initial condition at the horizon, the equation is 
integrated numerically
toward the $AdS_5$ boundary. In general, we find a linear combination
of normalizable and non-normalizable modes near the boundary. 
Since the new phase of the system should be represented by a normalizable
solution, we tune the initial condition at the horizon so that the
non-normalizable component vanishes. 
Figure \ref{fig:ads5Eh} describes numerical solutions for $\tau=0.35$.
The left graph shows the amplitude $h(r_+)=h_0$ at the horizon 
as a function of the momentum $k$.

\begin{figure}[ht]
\centering
\includegraphics[width=7cm]{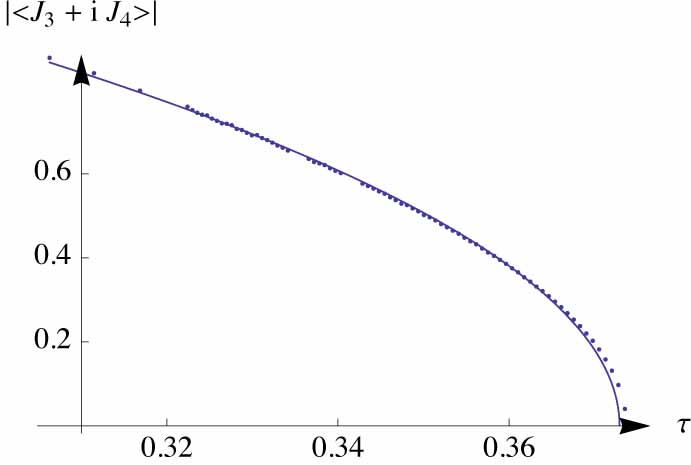}
\caption{The expectation value of the order parameter as a 
function of the temperature. The dotted curve is the numerical result
and the solid curve is its fit with $(1-\tau/\tau_c)^{1/2}$.}\label{fig:TvsO}
\end{figure}

Since there is a family of solutions parametrized by the momentum $k$, 
we need to choose the minimum energy density solution as the final state.
The energy density is given as a sum of the electric and magnetic energy
\begin{equation}
\mathcal H=\frac 1 {2\alpha^2} \int dr \left[ f'(r)^2 + \frac{k^2}{r^4} h(r)^2 + \frac{H(r)}{r^2} h'(r)^2\right]\;.
\end{equation}
It is convenient to rescale the energy density as 
$\tilde{ \mathcal H} = \alpha^2 r_+^4 \mathcal H$,
which is finite in the limit of $\alpha \rightarrow \infty$.
The right graph of Figure \ref{fig:ads5Eh} shows the energy density 
as a function of $k$.
Choosing the momentum corresponding to the minimum of the energy density, 
the expectation value of the order parameter $|\left<J_3 + i J_4\right>|$ can be read off from 
the asymptotic behavior of the corresponding bulk field $h(r)$.
Figure \ref{fig:TvsO} shows the expectation value $|\left<J_3 + i J_4\right>|$  
as a function of the rescaled temperature $\tau$. Near the critical temperature, it behaves as
\begin{equation}
|\left<J_3 + i J_4\right>| = A \left(1-\frac{\tau}{\tau_c}\right)^{\frac 1 2}\;,
\end{equation}
where $A=2.04$ and $\tau_c=0.37$.
The critical exponent $1/2$ is  typical for a mean field theory.
Indeed, this can be expected from the absence of quadratic terms in the equations of motion \eqref{E:EOMinRNAdSBH} 
and the fact that we consider the gravity system classically.
The mean field behavior is also observed in the holographic description of superconductivity \cite{Hartnoll:2008vx,Hartnoll:2008kx}.

\section{Quantum Corrections}\label{sec:corrections}

We found that the phase transition is of the second order in the classical supergravity approximation. 
In this section, we provide evidence that quantum corrections in the bulk change it to
the first order. 
Such a phenomenon has been observed by Brazovskii \cite{Brazovskii}
and elaborated by Swift and Hohenberg in \cite{SwiftHohenberg},
in the context of a classical statistical model at finite temperature.
We will extend this result to the gravity theory in $AdS_5$.

Let us review the Brazovskii model. 
It is a classical field theory in $d$ space dimensions
at finite temperature with 
the following scalar field Hamiltonian in the momentum representation,
\begin{equation}
\begin{split}
\mathcal{F}= &\frac 1 2 \int \frac{d^d q}{(2\pi)^d} \phi(\vec q)\phi(-\vec q) \left[ m^2+ (|\vec{q}|-q_0)^2\right]\\
&+ \frac{\lambda}{4!}\int \frac{d^d q_1}{(2\pi)^d}\frac{d^d q_2}{(2\pi)^d}\frac{d^d q_3}{(2\pi)^d} \phi(\vec q_1)\phi(\vec q_2)\phi(\vec q_3) \phi(-\vec q_1-\vec q_2 - \vec q_3)\;.
\end{split}
\end{equation}
Note the unconventional dispersion relation $\omega=m^2+ (|\vec q|-q_0)^2$.
It has the Van Hove singularity $d\omega/dq=0$ at $|\vec{q}|=q_0$, 
which will play an important role in the following. 

In the mean field approximation, 
the system undergoes a phase transition at $m^2 = 0$, and
there is a spatially modulated phase for $m^2 < 0$. 
The phase is characterized by the position dependent expectation value of $\phi(\vec x)$,
\begin{equation}\label{E:phivev}
\left< \phi(\vec x)\right> = 2 a \cos \vec q \cdot \vec x\;,
\end{equation}
with $|\vec q | = q_0$.
The phase transition is of the second order in this approximation.

\begin{figure}[ht]
\centering
\includegraphics[width=7cm]{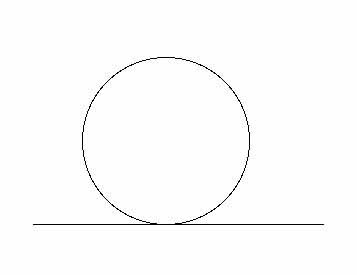}
\caption{A loop diagram that contributes to the two point correlation function.}
\label{fig:HartreeLoop}
\end{figure}

Let us examine if this picture is modified by thermal fluctuations. 
The inverse susceptibility is defined by 
\begin{equation}
M^2 := G^{-1}(|\vec q|=q_0)=m^2 - \Sigma(|\vec q|=q_0)\;,
\end{equation}
where $\Sigma$ represents thermal
loop contributions. In the Hartree approximation,
\begin{equation}\label{E:rcorrected}
M^2 = m^2 + \frac{s \lambda}{M} + \lambda a^2\;,
\end{equation}
where $s= \pi S_d q_0^{d-1}/(2\pi)^d$ with $S_d$ the area of the $(d-1)$-sphere.
There are higher order corrections, but they will not affect the behavior of $M^2$ near $a=0$.
The first term in the right-hand side
of \eqref{E:rcorrected} is the tree level value.
The second term comes from a loop diagram as
in Figure \ref{fig:HartreeLoop}, which gives a contribution near 
$q=q_0$ of the form
\begin{equation}\label{E:HartreeLoopContribution}
\frac{\lambda}{(2\pi)^d} 
\int  \frac{d^d \vec{q}}{M^2+(|\vec{q}|-q_0)^2} \sim 
\frac{\lambda}{(2\pi)^d} S_d q_0^{d-1}
\int_0^\infty 
\frac{dq}{M^2+(q-q_0)^2} = 
\frac{\lambda}{(2\pi)^d} \frac{\pi S_d q_0^{d-1}}{M} \;.
\end{equation}
There are subleading terms for small $M$, but 
what is relevant for the analysis below is the $1/M$ pole from the loop.
The third term $\lambda a^2$ comes from the quartic coupling combined with the
expectation value (\ref{E:phivev}) of $\phi$.

The free energy $\mathcal{F}$ for the expectation value
\eqref{E:phivev} can be evaluated by setting \cite{Brazovskii},
\begin{equation}\label{E:dFda}
\frac{d\mathcal{F}}{da} = 2 M^2 a - \lambda a^3\;.
\end{equation}
Combining this with 
\begin{equation}
\frac{da}{dM} = \frac{M}{\lambda a}\left(1+ \frac{s \lambda}{2 M^3}\right)\;,
\end{equation}
derived from  \eqref{E:rcorrected}, we find
\begin{equation}\label{E:FreeEnergyExpression}
\begin{split}
\mathcal{F} &= \int \frac{d\mathcal{F}}{da} da = \int \frac{d\mathcal{F}}{da} \frac{da}{dM} dM\\
&=\frac{1}{2\lambda}\left(\frac{M^4} 2 + m^2 M^2 + 3s \lambda M - \frac{s\lambda m^2}{M} -\frac{s^2\lambda^2}{2 M^2}\right)\;.
\end{split}
\end{equation}

\begin{figure}[ht]
\centering
\subfigure{
\includegraphics[width=5cm]{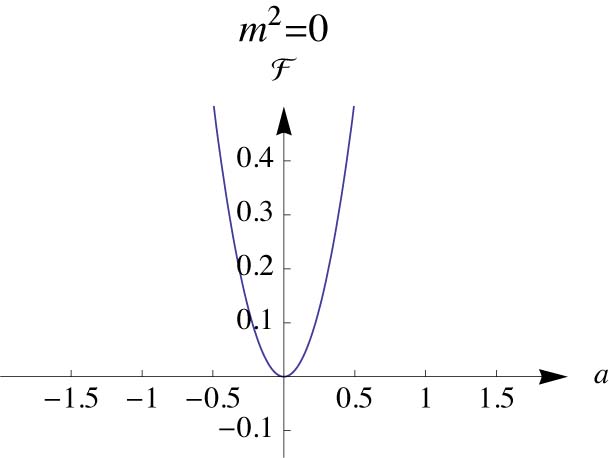}
}
\centering
\subfigure{
\includegraphics[width=5cm]{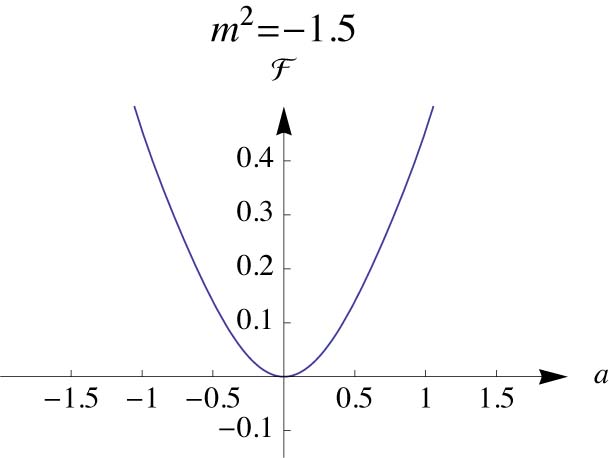}
}
\centering
\subfigure{
\includegraphics[width=5cm]{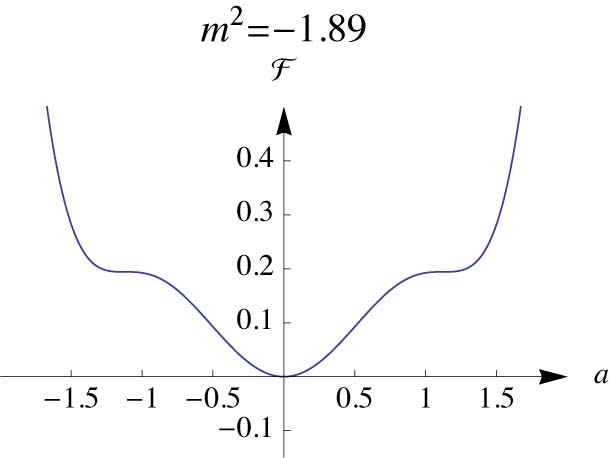}
}
\centering
\subfigure{
\includegraphics[width=5cm]{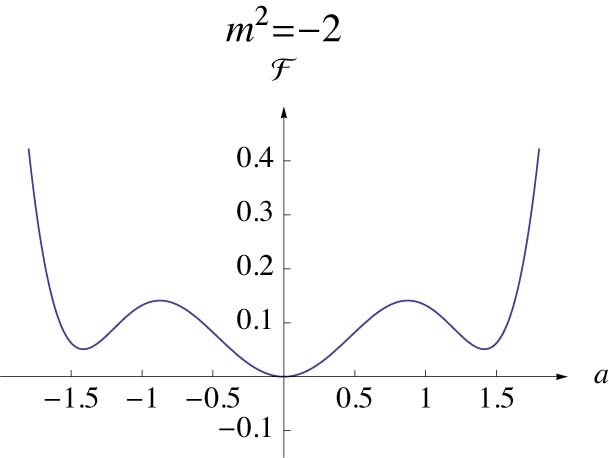}
}
\centering
\subfigure{
\includegraphics[width=5cm]{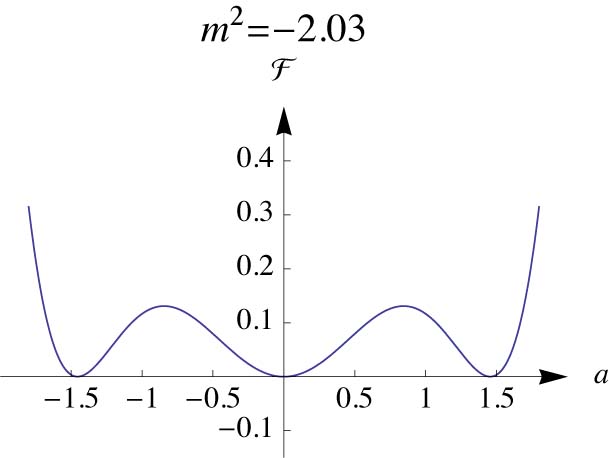}
}
\centering
\subfigure{
\includegraphics[width=5cm]{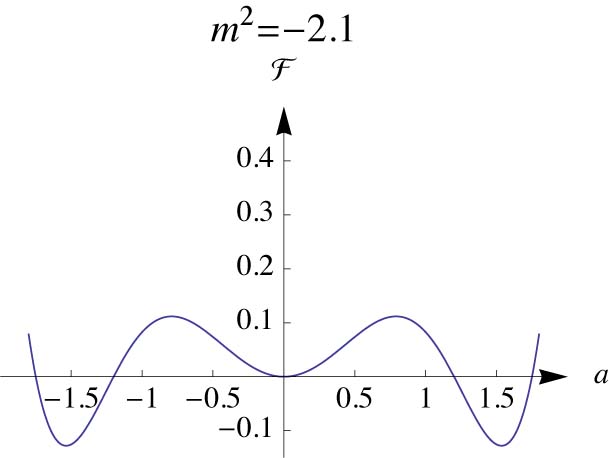}
}
\caption{Different shapes of the graphs of the free energy $\mathcal{F}$ as a function of $a$ as the parameter $m^2$ changes.}
\label{fig:Brazovskii}
\end{figure}

Note that the terms in the parenthesis in \eqref{E:FreeEnergyExpression} depend on $s$ and $\lambda$ only in the combination 
$s \lambda$. If we rescale $m$ and $M$ by a factor of $\epsilon$ and $s\lambda$ by $\epsilon^3$, the terms in the parenthesis 
in \eqref{E:FreeEnergyExpression} scales like $\epsilon^4$.
Thus, we can set $s=\lambda=1$ without loss of generality if we do not care about an overall factor of $\mathcal{F}$.
Figure \ref{fig:Brazovskii} illustrates the free energy $\mathcal{F}$ as a function of $m^2$.
For $m^2\geq 0$, we have a graph with only a minimum at the origin.
As $m^2$ decreases and crosses $-2^{-\frac 2 3} 3 \sim -1.89$, 
we see two inflection points away from the origin,
representing the spatially modulated phase. At this point, the homogeneous 
phase at $a=0$ still has a lower energy.
As we lower $m^2$ further, the energy of the spatially modulated phase 
becomes lower than that of the homogeneous phase, and
 the first order phase transition occurs.

\begin{figure}[ht]
\centering
\includegraphics[width=7cm]{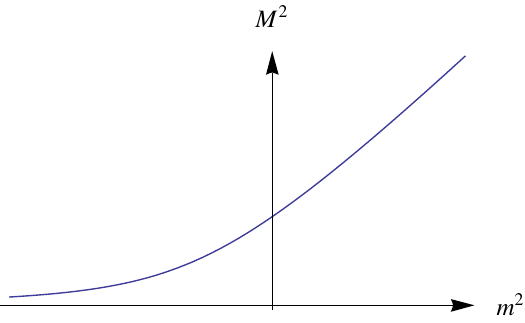}
\caption{The relation between $m^2$ and $M^2$. 
 Note that for any value of $m^2$, there is 
always a solution with $M^2>0$.}
\label{fig:m2M2}
\end{figure}

The crucial point is that the origin $a=0$ remains semi-stable
throughout the process since $d^2 \mathcal F / da^2 =2M^2 >0$ at $a=0$.
This in turn is due to the fact that
\eqref{E:rcorrected} has a solution with $M^2 > 0$
for any value of $m^2$ including $m^2 < 0$. This is possible 
since the second term in the equation is
proportional to $1/M$ and diverges as $M\rightarrow 0$. This is 
illustrated in Figure \ref{fig:m2M2}.
Since $a=0$ is always semi-stable, 
the phase transition cannot be of the second order.

The $1/M$ singularity in \eqref{E:rcorrected} originates
from  the fact that
the kinetic operator
$\left[M^2+(|\vec q|-q_0)^2\right]$ vanishes at $M^2=0$ and $|\vec{q}| = q_0$. 
If the zero of the kinetic operator were at $\vec{q}=0$, 
it would not have caused the $1/M$ singularity because the 
factor $q^{d-1}$ in the phase space volume $q^{d-1}dq$ would 
have suppressed it. The singularity is generated in
our case because of the larger phase space volume $\sim q_0^{d-1}dq$
at $|\vec{q}|=q_0$ in \eqref{E:HartreeLoopContribution}. It ensures that there is a positive 
$M^2$ solution to \eqref{E:rcorrected} and that 
the origin of the field space is meta-stable.

This feature is shared by the gravity theory considered in this paper. 
In our classical analysis in the last section, we found that
the phase transition happens at $T=T_c$
and the new phase for $T < T_c$ is represented by the non-linear 
static solutions constructed in section \ref{sec:MCSSBHAdS}.
Since the phase transition is of the second order, 
the size of the non-linear solution grows linearly in  
$\sqrt{T_c - T}$. This means that, at $T=T_c$, there are static 
solutions to the linearized equations of motion, namely a
zero of the kinetic operator at non-zero momenta. 
As in the case of the Brazovskii model, 
it  generates a $1/M$ singularity in the
two-point function.  Thus, we expect that
the homogeneous phase will remain meta-stable for $T < T_c$.
If quantum corrections are parametrically suppressed 
(e.g., by $1/N$),  the spatially modulated phase will eventually acquire
lower energy and the first order phase transition will take place
at that point.

\section{Linear response}\label{sec:GreensFunction}

Finally, let us examine linear response of our solution when we couple a gauge field to the current $J$ at the boundary.
Because of the Chern-Simons term, the current is anomalous. We therefore treat the boundary gauge field as an external and
non-dynamical source as in recent papers, for example \cite{Son, D'Hoker}. 

Note that the background solution we consider is inhomogeneous, carrying the momentum $k$. 
In the homogeneous setup, it is natural to choose a translationally invariant source.
In our case, we consider a small perturbation of the form $\tilde A=\tilde A^{(0)}+a$, 
where $\tilde A^{(0)}$ is our non-linear solution and $a$ is a small perturbation with non-zero components, 
\begin{equation}\label{E:aAnsatz}
a_3(x_2,r,t)+i a_4(x_2,r,t)= - i a_T(r,t) e^{-i k x_2},~~~ a_2(r,t)=a_L(r,t)\;.
\end{equation}
Here, $a_T(r,t)$ and $a_L(r,t)$ are real functions of $r$ and $t$.
Notice that we turn on a magnetic field as well as an electric field.
It is not possible to turn on only an electric field due to the Bianchi identity $\partial_{[+} F_{t2]}=0$, where $+$ denotes the coordinate $x_3+i x_4$.
However, in the current setup, the magnetic field is determined by the electric field and is not an independent quantity.

The non-trivial linear equations of motion of the fields $a_T(r,t)$ and $a_L(r,t)$ are
\begin{equation}\label{E:CoupledEquationsForConductivity}
\begin{split}
-\frac{r}{H(r)} \partial_t^2 a_L + \partial_r( r H(r) \partial_r a_L) + 4 \partial_r h(r) \partial_t a_T &=0\\ 
-\frac{r}{H(r)} \partial_t^2 a_T +\partial_r(r H(r) \partial_r a_T) -\frac{k^2} r a_T + 4 k \frac{\tilde E-2\alpha k h(r)^2}{r^3} a_T -4 \partial_r h(r) \partial_t a_L& =0\;.
\end{split}
\end{equation}
Here $h(r)$ is the non-linear solution we have found previously.
We are interested in modes with definite frequency of the form $e^{-i \omega t}$.
The fields $a_T(r)$ and $a_L(r)$ behave as $H(r)^{-\frac{i \omega} 4}$ near the horizon $r=r_+$ if we impose the in-going boundary condition.
On the other hand, near the $AdS_5$ boundary $r=\infty$, they behave as
\begin{equation}\label{E:aa2nearAdS5}
\begin{split}
a_T&= a_{T(0)} + a_{T(2)} r^{-2} + a_{T(\log)} r^{-2} \log r/r_+ +\cdots\\
a_L&= a_{L(0)} + a_{L(2)} r^{-2} + a_{L(\log)} r^{-2} \log r/r_+ +\cdots\;.
\end{split}
\end{equation}
Note that there are logarithmic terms and that
the coefficient $a_{T(2)}$ or $a_{L(2)}$ can be shifted by $a_{T(\log)}$ and $a_{L(\log)}$ if we change the scale of 
the radial coordinate $r$.
This corresponds to the choice of a renormalization scale, and should not affect physical quantities.
The behavior of the equations \eqref{E:CoupledEquationsForConductivity} near the $AdS_5$ boundary $r=\infty$ 
fixes $a_{T(\log)}$ and $a_{L(\log)}$ to be
\begin{equation}\label{E:logTerms}
a_{T(\log)}=\frac 1 2 (\omega^2 - k^2) a_{T(0)}\;,\qquad a_{L(\log)} = \frac{\omega^2}{2} a_{L(0)}\;.
\end{equation}
Since we are solving a set of linear differential equations, the coefficients $a_{T(2)}$ and $a_{L(2)}$ are determined linearly from $a_{T(0)}$ and $a_{L(0)}$. That is,
\begin{equation}\label{E:omega}
\begin{pmatrix} a_{T(2)} \\ a_{L(2)} \end{pmatrix} = \Omega  \begin{pmatrix} a_{T(0)} \\ a_{L(0)} \end{pmatrix}\;,
\end{equation}
for some $2\times 2$ complex matrix $\Omega$.
From the prescription in \cite{Son:2002sd}, and analogously to \cite{Horowitz:2008bn}, the retarded Green's functions for fields $a_T$ and $a_{L}$ are given by
\begin{equation}
G^R = 2 \Omega + \begin{pmatrix} \left(\omega^2-k^2\right)\left( \log \frac{r}{r_+}- \frac 1 2 \right) & 0 \\ 0 & \omega^2\left(\log \frac{r}{r_+}- \frac 1 2\right)\end{pmatrix}\;,
\end{equation}
where the last term is due to the logarithmically divergent terms in \eqref{E:aa2nearAdS5} whose coefficients are given in \eqref{E:logTerms}.
These divergent parts can be removed by adding a suitable boundary counterterm to the gravity action.
After subtracting the divergence, we arrive at
\begin{equation}
G^R = 2 \Omega - \begin{pmatrix} \frac 1 2 \left(\omega^2-k^2\right) & 0 \\ 0 & \frac 1 2\omega^2\end{pmatrix}\;,
\end{equation}
When considering modes of definite frequency, the additional electric fields $E_T$ and $E_L$ are $i \omega a_{T(0)}$ and $i \omega a_{L(0)}$, respectively.
Therefore the conductivity is given by
\begin{equation}
\sigma = \frac{1}{i\omega}G^R = \frac{2}{i\omega} \Omega + \begin{pmatrix} \frac{i}{2\omega}\left(\omega^2-k^2\right) & 0 \\ 0 & \frac{i \omega}2\end{pmatrix}\;.
\end{equation}

Let us evaluate the conductivity $\sigma$ numerically for the Chern-Simons coupling $\alpha=0.59$.
For this value of $\alpha$, the minimum energy occurs at the momentum $k=-2.37 r_+$.
We denote the components of the retarded Green's function and the conductivity respectively as
\begin{equation}
G^R=\begin{pmatrix} G_{TT} & G_{TL}\\ G_{LT} & G_{LL} \end{pmatrix}\;,
\end{equation}
and
\begin{equation}
\sigma=\begin{pmatrix} \sigma_{TT} & \sigma_{TL}\\ \sigma_{LT} & \sigma_{LL} \end{pmatrix}\;.
\end{equation}

\begin{figure}[ht]
	\begin{center}
		\subfigure{\includegraphics[height=3.5cm]{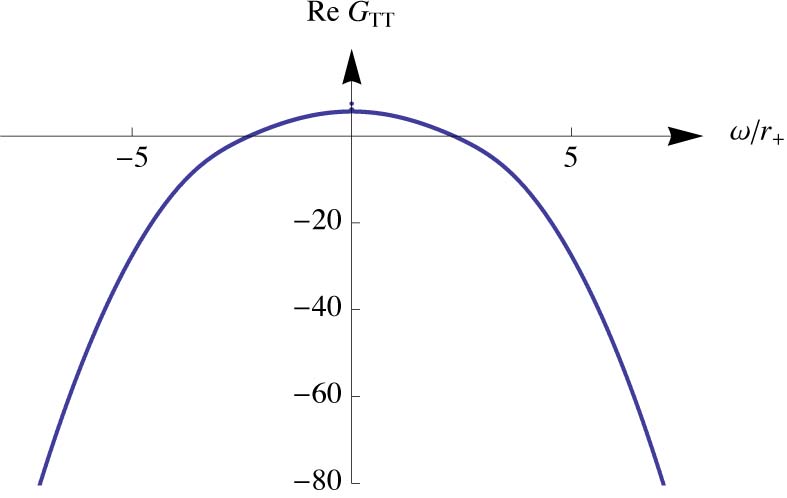}}
		\subfigure{\includegraphics[height=3.5cm]{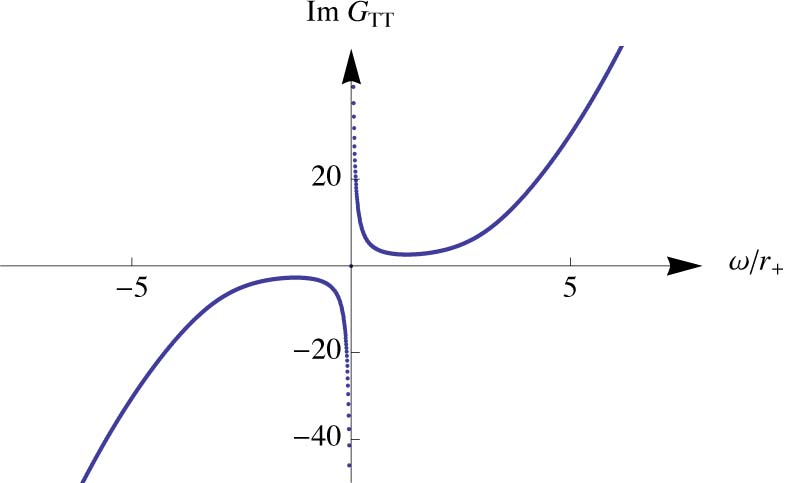}}
	\end{center}
	\caption{The real and the imaginary parts of $G^R_{TT}$.}
	\label{fig:G11}
\end{figure}

\begin{figure}[ht]
		\begin{center}
			\subfigure{\includegraphics[height=3.5cm]{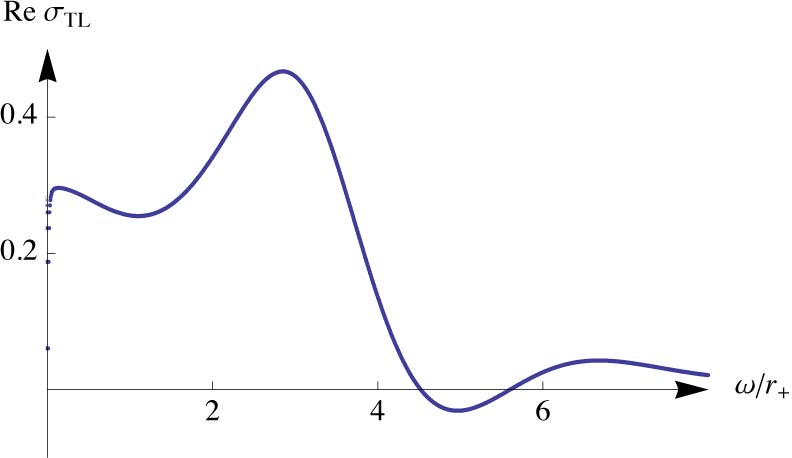}}
			\subfigure{\includegraphics[height=3.5cm]{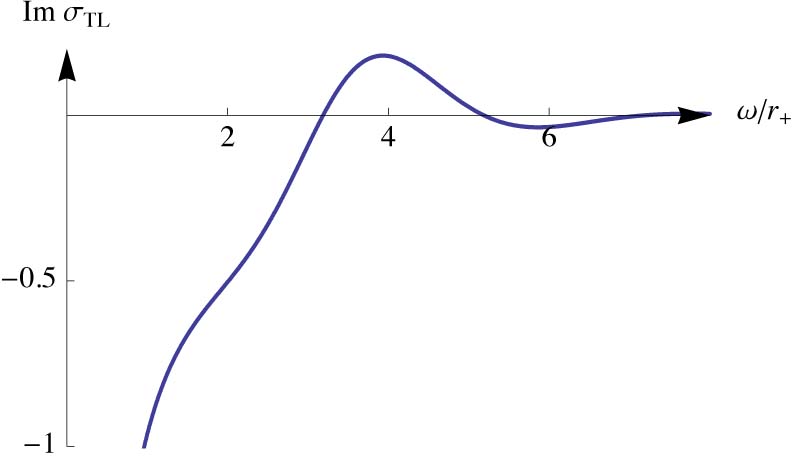}}
		\end{center}
		\begin{center}
			\subfigure{\includegraphics[height=3.5cm]{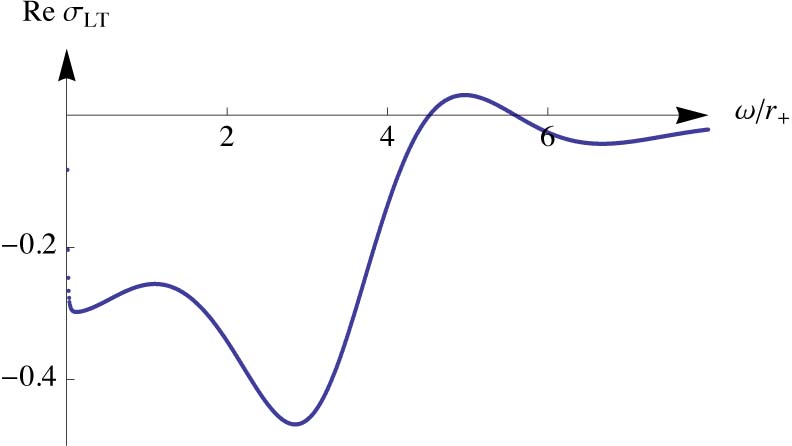}}
			\subfigure{\includegraphics[height=3.5cm]{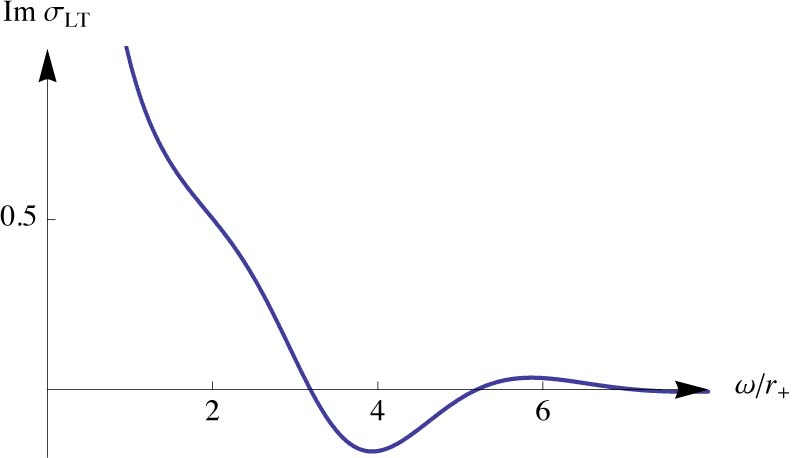}}
		\end{center}
\caption{The real and the imaginary parts of $\sigma_{TL}$ and $\sigma_{LT}$.}
\label{fig:sigma12}
\end{figure}

This numerical result has several interesting features.
The imaginary part of the retarded Green's function 
$G_{TT}$ for a pair of $a_T$ fields behaves like $1/\omega$ 
near $\omega=0$, as shown in Figure \ref{fig:G11}.
The pole at $\omega=0$ is directly related to the fact 
that $a_T$ is a Goldstone mode.
This can be seen as follows.
As $\omega$ goes to 0, the solution becomes static and we can ignore 
the terms with time derivatives 
in \eqref{E:CoupledEquationsForConductivity}.
In this limit, the equation of motion for $a_T$ reduces to 
the linearized version of the equation for the phase 
rotation of the inhomogeneous background solution 
\eqref{E:inhomogeneousAnsatz}.
Especially, $a_T$ goes to $h(r)$ as we take a static limit.
Since $h(r)$ is a solution that does not have a non-normalizable mode, if we start with a source such that $a_{T(0)}=0$, \eqref{E:omega} implies that $\Omega_{TT}$ should diverge in the static limit.
Therefore, we expect to see a pole at $\omega=0$ in $G^R_{TT}$.

Figure \ref{fig:sigma12} shows the off-diagonal components of the conductivity $\mathrm{Im} \, \sigma_{TL}$ and $\mathrm{Im}\, \sigma_{LT}$.
Note that they behave as $1/\omega$ with opposite coefficients.
The antisymmetry of $\sigma_{TL}$ and $\sigma_{LT}$ is due to the fact that
\eqref{E:CoupledEquationsForConductivity} remains invariant under the change $t\rightarrow -t$ if we simultaneously change the sign of $a_T(r)$.
The poles at $\omega=0$ in the imaginary part of the conductivity indicate that there are delta function contributions at $\omega=0$ to $\mathrm{Re}\, \sigma_{TL}$ and $\mathrm{Re} \,\sigma_{LT}$, which implies that there is an off-diagonal infinite DC conductivity.

\begin{figure}[ht]
\centering
\includegraphics[width=7cm]{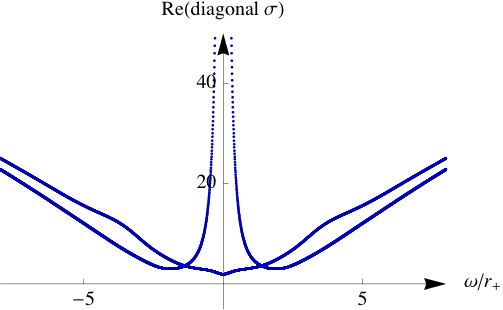}
\caption{The real parts of the diagonal components of the diagonalized conductivity matrix.}
\label{fig:DiagonalizedConductivities}
\end{figure}

It is also instructive to diagonalize the conductivity matrix.
Figure \ref{fig:DiagonalizedConductivities} shows the real parts of the two diagonal components of the diagonalized conductivity matrix.
The real part of a conductivity is directly related to the spectral density, and
the real parts of the two eigenvalues measure the spectral densities corresponding to two linear combinations of the transverse and longitudinal currents associated with $a_T$ and $a_L$ fields.
In particular it should always be positive, and
indeed this is shown explicitly in Figure \ref{fig:DiagonalizedConductivities}.
One of the diagonal conductivities has a $1/\omega^2$ pole, which is again related to the presence of a Goldstone mode.

\section*{Acknowledgments}

We thank Michael Cross, Per Kraus, Shin Nakamura and Dam T. Son 
for discussions. We also thank Sean Hartnoll and Subir Sachdev 
for their comments on the earlier version of this paper.  
We are grateful to Hermann Nicolai and to the Max-Planck-Institut
f\"ur Gravitationsphysik for hospitality.
C.P. thanks the hospitality of the Korea Institute for Advanced Study and 
the Institute for the Physics and Mathematics of the Universe at the University of Tokyo.
H.O. thanks the Aspen Center for Physics, where this work
was completed, for the hospitality. 
 
This work is supported in part by the DOE grant DE-FG03-92-ER40701
and the World Premier International Research Center Initiative of 
MEXT. H.~O. is also supported in part by
JSPS Grant-in-Aid for Scientific Research (C) 20540256
and by the Humboldt Research Award.

\begin{appendix}

\section{Analysis in the $AdS_2\times \mathbb{R}^3$ geometry}\label{sec:MCSAdS2R3}

In this appendix, we consider the Maxwell field in 
the $AdS_2\times \mathbb{R}^3$ geometry ignoring its back-reaction 
to the metric. Since the geometry corresponds to the zero temperature
limit of the black hole, it does not appear in the limit we consider in
section 3. On the other hand, 
the equations of motion can be solved analytically in this case,
and it gives a natural generalization of the analysis in section 2.  

The metric of $AdS_2\times \mathbb{R}^3$ is given by
\begin{equation}
ds^2= - r^2 dt^2 + \frac{dr^2}{r^2} + dx_2^2+ dx_3^2 + dx_4^2\;.
\end{equation}
Note that, especially, $\sqrt{-g}=1$.
We assume the radius of curvature for $AdS_2$ is 1.
A different value of the radius can be easily considered by using dimensional analysis. For a generic metric, the Lagrangian is given by
\begin{equation}
\mathcal{L} = -\frac 1 4 F_{IJ}F^{IJ} + \frac{\alpha}{3!\sqrt{-g}} \epsilon^{IJKLM} A_I F_{JK} F_{LM}\;,
\end{equation}
whose equation of motion is
\begin{equation}
\partial_J (\sqrt{-g} F^{JI}) + \frac{\alpha}{2} \epsilon^{IJKLM} F_{JK}F_{LM} =0\;.
\end{equation}
The index is such that $t=0$ and $r=1$. Here $\sqrt{-g}=1$.

Let us find out a class of non-linear static solutions with the ansatz
\begin{equation}
\begin{split}
&A_0 = f(r)\;,\qquad A_1=A_2=0\\
&A:=A_3 +i A_4 = h(r) e^{-i k x_2}\;.
\end{split}
\end{equation} 
The equations of motion can be written as
\begin{equation}\label{E:AdS2R3EOM}
\begin{split}
-f''(r) - 4\alpha k h(r) h'(r) &= 0\\
\partial_r (r^2 h'(r)) - k^2 h(r) + 4\alpha k h(r) f'(r) &=0\;.
\end{split}
\end{equation}
Integrating the first equation, we obtain
\begin{equation}
f'(r) + 2\alpha k h(r)^2 = E
\end{equation}
where $E$ is the constant electric background field since the left hand side is just the first component of the conjugate momentum $\Pi_1$.
Solving for $f'(r)$ and plugging into \eqref{E:AdS2R3EOM},
\begin{equation}
\partial_r ( r^2 h'(r)) - k^2 h(r) + 4\alpha k h(r) (E- 2\alpha k h(r)^2 ) =0\;.
\end{equation}
For $u=\log r$, the equation becomes
\begin{equation}
h''(u) + h'(u) - k^2 h(u) + 4\alpha k h(u) (E- 2\alpha k h(u)^2)=0\;,
\end{equation}
If we treat $u$ as a time coordinate, this equation describes a one-dimensional motion of a particle parametrized by $h$ subject to a frictional force and under the potential
\begin{equation}
U= \frac 1 2 k (4\alpha E - k) h^2  - 2\alpha^2 k^2 h^4\;.
\end{equation}
If $0<k<4\alpha E$, the potential is an upside-down Mexican hat. 
If a particle starts at one of the two hills at $u=-\infty$, 
it will oscillate around $h=0$, as in the case of the Minkowski 
space discussed in section 2. The new feature in the $AdS_2$ case is
that there is a friction term in (A.8) and the motion will 
eventually stop at $h=0$.

In order for a non-linear solution to exist, the momentum $k$ must obey
the additional condition, $k(k-4\alpha E )>-\frac 1 4$,
which is equivalent to the Breitenlohner-Freedman bound.
This condition is needed since any non-linear solution $h$ tends to 
$h=0$ for large $u = \log r$ and obeys the linearized equation near
the boundary of $AdS_2$. This in particular means that there is no
non-linear solution with $k=0$. Unlike the case of the Minkowski space,
the end-point of the instability is not a trivial vacuum but a
non-trivial solution carrying a non-zero momentum. 

\end{appendix}

\end{document}